\documentclass[12pt,twoside]{article}   
\usepackage[super,sort,comma]{natbib}

\usepackage{fancyhdr}		

\usepackage[section]{placeins}   %

\usepackage{graphicx}

\usepackage{amsmath}

\usepackage{soul} 
\usepackage{comment}
\usepackage{changepage}

\makeatletter \renewcommand\@biblabel[1]{$^{#1}$} \makeatother
\newcommand{\cen}[1]{\begin{center} #1 \end{center}}

\newcommand{\pz}{\phantom{0}}

\usepackage[mathlines]{lineno}

\usepackage{hyperref}
\hypersetup{ colorlinks,
    citecolor=blue,
    filecolor=blue,
    linkcolor=blue,
    urlcolor=blue
}

\usepackage{xcolor}

\definecolor{gray}{rgb}{0.6,0.6,0.6}
\definecolor{red}{rgb}{0.85,0,0}
\definecolor{green}{rgb}{0,0.85,0}
\definecolor{blue}{rgb}{0,0,0.85}
\definecolor{beige}{rgb}{0.92,0.87,0.78}
\definecolor{lightgreen}{rgb}{0.5,0.99,0.5}
\definecolor{lightcyan}{rgb}{0.5,0.99,0.99}
\definecolor{lightred}{rgb}{0.99,0.8,0.8}

\usepackage[all]{hypcap}    

\begin{document}

\cen{\sf {\Large {\bfseries A Comparison of Proton Stopping Power Measured with Proton CT and x-ray CT in Fresh Post-Mortem Porcine Structures} \\
\vspace*{10mm}
Don F. DeJongh$^1$, 
Ethan A. DeJongh$^1$, 
Victor Rykalin$^1$, 
Greg DeFillippo$^2$, 
Mark Pankuch$^2$, 
Andrew W. Best$^3$, 
George Coutrakon$^3$, 
Kirk L. Duffin$^4$, 
Nicholas T. Karonis$^{4,5}$, 
Caesar E. Ordo\~{n}ez$^4$, 
Christina Sarosiek$^3$, 
Reinhard W. Schulte$^6$, 
John R. Winans$^4$, 
Alec M. Block$^{7,8}$, 
Courtney L. Hentz$^{7,8}$,
James S. Welsh$^{7,8}$
}
\vspace{5mm}\\
$^1$ProtonVDA LLC, Naperville, IL 60563, USA\\
$^2$Northwestern Medicine Chicago Proton Center, Warrenville, IL 60555, USA\\
$^3$Department of Physics, Northern Illinois University, DeKalb, IL 60115, USA\\
$^4$Department of Computer Science, Northern Illinois University, DeKalb, IL 60115, USA\\
$^5$Argonne National Laboratory, Data Science and Learning Division, Argonne, IL 60439, USA\\
$^6$Loma Linda University, Loma Linda, CA 92350, USA\\
$^7$Edward Hines Jr VA Medical Center, Radiation Oncology Service, Hines, IL 60141, USA\\
$^8$Department of Radiation Oncology, Loyola University Stritch School of Medicine, Maywood, IL 60153, USA\\
\vspace{5mm}
}

\pagenumbering{roman}
\setcounter{page}{1}
\pagestyle{plain}
\cen{$^+$Author to whom correspondence should be addressed. email: fritz.dejongh@protonvda.com}

\begin{abstract}
\noindent {\bf Purpose:} Currently, calculations of proton range in proton therapy patients are based on a conversion of CT Hounsfield Units of patient tissues into proton relative stopping power. Uncertainties in this conversion necessitate larger proximal and distal planned target volume margins. Proton CT can potentially reduce these uncertainties by directly measuring proton stopping power. We aim to demonstrate proton CT imaging with complex porcine samples, to analyze in detail three-dimensional regions of interest, and to compare proton stopping powers directly measured by proton CT to those determined from x-ray CT scans. \\
{\bf Methods:} We have used a prototype proton imaging system with single proton tracking to acquire proton radiography and proton CT images of a sample of porcine pectoral girdle and ribs, and a pig's head.  We also acquired close in time x-ray CT scans of the same samples, and compared proton stopping power measurements from the two modalities. In the case of the pig's head, we obtained x-ray CT scans from two different scanners, and compared results from high-dose and low-dose settings. \\
{\bf Results:} Comparing our reconstructed proton CT images with images derived from x-ray CT scans, we find agreement within 1\% to 2\% for soft tissues, and discrepancies of up to 6\% for compact bone. We also observed large discrepancies, up to 40\%, for cavitated regions with mixed content of air, soft tissue, and bone, such as sinus cavities or tympanic bullae. \\
{\bf Conclusions:} Our images and findings from a clinically realistic proton CT scanner demonstrate the potential for proton CT to be used for low-dose treatment planning with reduced margins.
\\

\end{abstract}

\newpage     



\setlength{\baselineskip}{0.1cm}      

\pagenumbering{arabic}
\setcounter{page}{1}
\pagestyle{fancy}

\section{Introduction}

In radiation therapy, protons provide different dose distributions compared to x-rays, with a relatively low dose deposition in the entrance region (plateau), followed by a steep increase to a dose (Bragg) peak and an even steeper distal dose fall-off. Treatment planning for proton therapy starts with an x-ray CT image with voxel values measured in Hounsfield units (HU), converted to proton relative stopping power (RSP), relative to water, using scanner-specific calibrations.  There is no consistent one-to-one relationship between HU and RSP for different tissues and materials, leading to significant range uncertainties, typically quantified as 3.5\% $\pm$ 1 mm for treatment planning purposes at most proton treatment centers \cite{Paganetti_2012}.

Treatment planning procedures take these uncertainties into account with measures including adding uncertainty margins, selection of beam angles tangential to organs at risk, and robust optimization.  Using dose delivery technology such as pencil beam scanning (PBS) and intensity modulated proton therapy, the resulting plans are robust to the uncertainties and provide major benefits to a significant fraction of patients \cite{Lomax}.  However, plans with larger margins increase the high-dose treatment volume and often preclude use of the most advantageous beam angles. In the quest to further optimize proton therapy, proton beam-based image guidance is often considered to be a prerequisite to achieve the full potential of proton therapy \cite{Schreuder}. This is particularly the case for hypo-fractionated treatments, which can benefit from more conformal dose distributions and range verification given the high dose delivery for each treatment.  

Proton CT (pCT) may substantially reduce the uncertainties of treatment planning by directly measuring RSP, and provides further benefits including lack of artifacts from metallic implants, and much lower dose to the patient than comparable x-ray images \cite{Schulte}.  
The ability to directly measure accurate RSP maps would also enable pCT to be used for cross-calibration of x-ray CT systems~\cite{Farace}.
Further, pCT enables proton radiography (pRad), which has the potential to provide a fast and efficient check of patient set up and integrated range along the beam’s eye view just before treatment \cite{Miller,Pankuch}. A comparison of a pRad with a digitally-reconstructed radiograph (DRR) based on projections through the 3D RSP map used for treatment planning can reveal discrepancies from range uncertainties as well as other sources such as changes in anatomical consistency and patient misalignment.

In a recent publication, Deffet \textit{et al.} have assessed the accuracy of HU to RSP conversion with a pRad of a pig's head \cite{https://doi.org/10.1002/mp.14571}.  Using a detector that integrates the contribution of every proton during the irradiation of a pencil beam shot, they applied a deconvolution algorithm to improve the spatial resolution to better than the pencil beam size, and registered the resulting pRad with an x-ray CT.  Comparing the pRad to the expectation from the CT, they find range uncertainties generally well below the typical safety margins by about a factor of five, with the exception of regions including sinus cavities, where the range discrepancies are larger.

Meijers \textit{et al.} have studied range uncertainties in porcine samples of head, thorax, and femur using a range probing technique~\cite{Meijers_2020}.  They found that the maximum discrepancy was 75\% of the uncertainty margin prescription.  A similar study of 2025 proton spots acquired during different breathing phases with a set of inflatable porcine lungs found a mean relative range error of 1.2\% with a spread of 2.3\% (1.5 standard deviations)~\cite{Meijers_2020_lungs}.

B\"{a}r \textit{et al.} experimentally compared dual-energy CT (DECT) performance against conventional single-energy CT (SECT) for proton treatment planning by comparing 12 tissue-equivalent plastic materials and 12 fresh heterogeneous animal tissue samples in terms of water equivalent range predictions of 195 MeV protons~\cite{baer-tissue}. While SECT performed well in soft tissues, they found larger deviations in bone, and better performance of DECT compared to SECT in all materials.

We present herein measurements of RSP in a fresh post-mortem pig's head as well as a sample of porcine pectoral girdle and ribs, to compare the RSP determination from pCT and x-ray CT images taken close in time.
The 3D image reconstructions enabled us to define regions of interest (ROI) and directly compare the voxelized RSP of different tissues in the porcine samples. We also present a difference map between a pRad image and an x-ray DRR.

\section{Materials and Methods}

\subsection{The proton imaging system}

We obtained proton images with the prototype clinical ProtonVDA proton imaging system~\cite{dejongh2020technical}, positioned in a horizontal beam line equipped with PBS at the Northwestern Medicine Chicago Proton Center (NMCPC). A rotating platform between the tracking planes enabled imaging with a full set of angles relative to the PBS system.
In contrast to Deffet {\it et al.}\cite{https://doi.org/10.1002/mp.14571}, we obtained proton images by measuring trajectories and residual ranges of individual protons. Tracking detectors measure the transverse positions of individual protons before and after the object, and a residual range detector determines the proton energy absorbed within the object.  Proton trajectories deviate from straight lines due to multiple Coulomb scattering, and the estimation of individual proton paths improves spatial resolution compared to systems that integrate charge from pencil beams. 

The ProtonVDA system is capable of automatically acquiring data from a low-intensity proton beam (a few million protons per second) delivered by the PBS system and reconstructing pRad images \cite{sarosiek2020prototype} with a software platform~\cite{Ordonez} that processes the data as they are received with no human intervention needed after the start of the beam delivery.  While we are in the process of extending this automatic capability to pCT, for the current work we  acquired data in 90 separate segments with a 4$^\circ$ rotation between segments to provide the full set of angles required for pCT, and processed the data offline. Our pRad system uses multiple incoming proton energies for a single image\cite{dejongh2020technical}, {and the advantages of this approach, including the use of a thin range detector, also apply to pCT}. We acquired data with multiple energies at each angle, with a scan pattern {consisting of spots separated by 0.5 cm and} filling a $25\times25$ cm$^2$ area for each energy at each angle. {Each scan pattern used 3.3 seconds of beam time, for 5 minutes total beam time per energy. The use of object-specific scan patterns would greatly optimize and reduce the required beam time, while keeping the residual range low across the field}\cite{dejongh2020technical}. {For the current study we selected offline the data with the appropriate energy for each region at each angle}.  For each proton, we calculated a water-equivalent path length (WEPL) using a calibration procedure that relates the response of our detector to the clinically validated range settings of the proton therapy system~\cite{dejongh2020technical}. For each proton, we use the incoming direction and tracking detector data to calculate the most likely path (MLP), with spatial resolution typically near 1 mm~\cite{sarosiek2020prototype}. We have previously verified our ability to measure accurate WEPL and estimate MLP for protons in the context of reconstructing pRad images\cite{sarosiek2020prototype}. 

We reconstructed pCT images as RSP maps, using a least-squares iterative algorithm~\cite{dejongh2020iterative} which provides assurance that the iterative solution has converged with a fit to the WEPL and MLP data that is close to optimal within the statistical power of the data. With this fit, the RSPs of the voxels along the trajectories of the protons on average account for the total WEPLs of those protons. We did not apply smoothing filters to reduce the statistical noise inherent in our data.

\subsection{Validation of pCT RSP accuracy}

We tested the accuracy of our pCT images with a custom cylindrical phantom which incorporates a set of eight 4 cm high, 1.8 cm diameter tissue-equivalent cylindrical inserts. Each insert was milled from a $10 \times 10 \times 2$ cm$^3$ slab of material.  Prior to the manufacturing of the inserts, the RSP of each slab was measured with a multi-layer ionization chamber used for range calibrations at NMCPC by observing the pull-back of a 150 MeV proton Bragg peak with and without the slabs in place\cite{sarosiek2020prototype}.
The pCT image of the custom cylindrical phantom used a total of approximately 20 million protons taken at incoming energies of 118, 160, and 187 MeV to reconstruct a volume of $200 \times 60 \times 200$ 1 mm$^3$ voxels.  

\subsection{pCT and x-ray CT image acquisition}

\begin{figure}
  \begin{center}
  \includegraphics[width=0.16\textwidth]{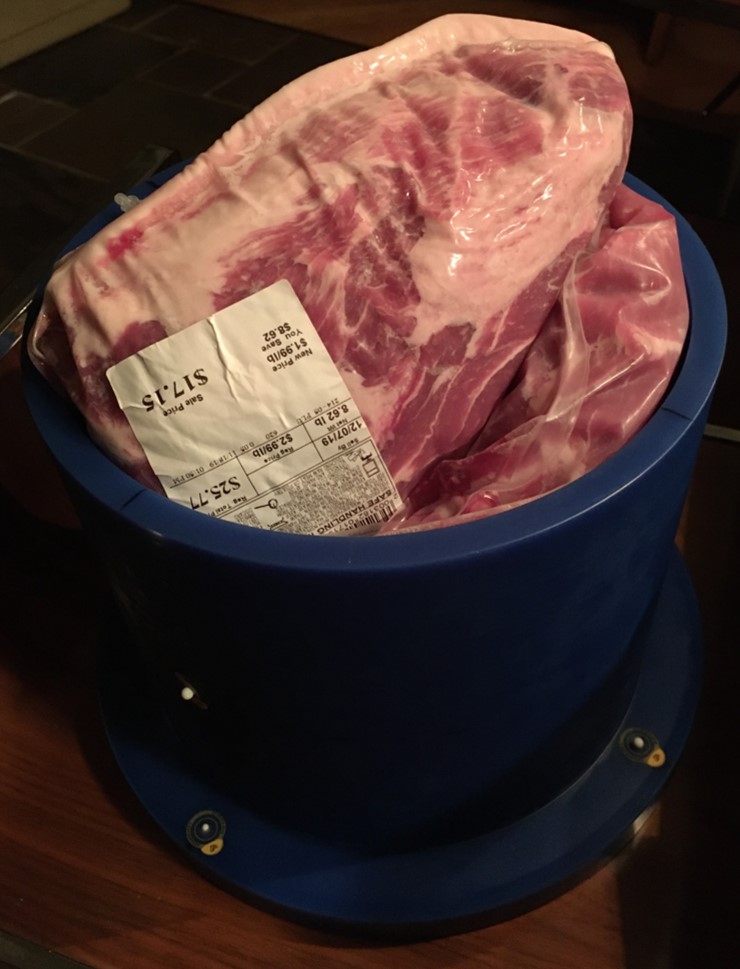}\includegraphics[width=0.28\textwidth]{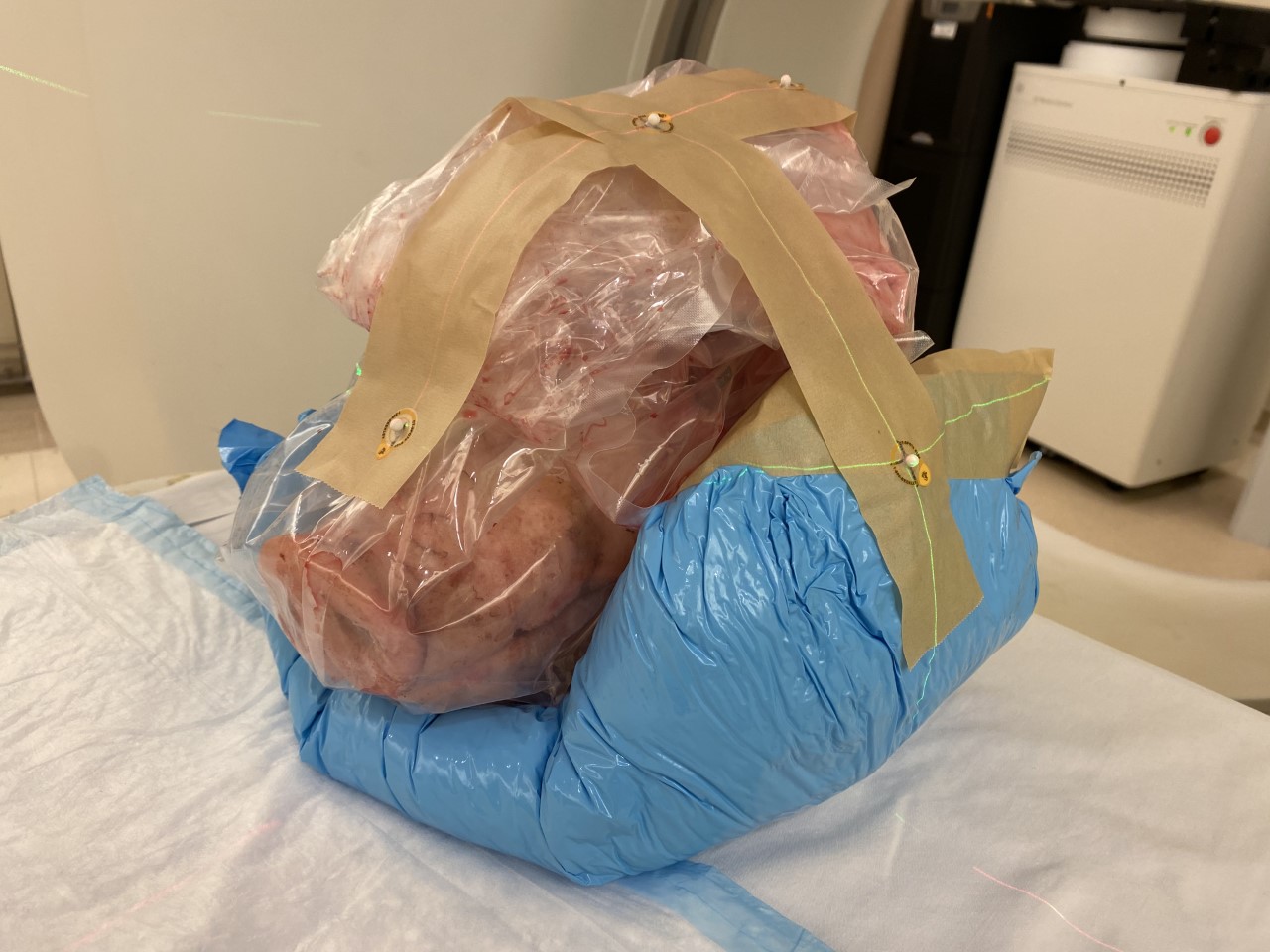}
  \caption{Left:  Porcine girdle and ribs, inserted into a blue-wax bucket.  Right:  Pig's head immobilized with plastic foam and tape, and with surface-mounted spherical CT markers.}\label{setup}
  \end{center}
\end{figure}

We assembled two fresh post-mortem porcine structures for proton imaging from samples available from a butcher's shop, shown in Fig.~\ref{setup}.
The first is a sample of porcine pectoral girdle (shoulder) and ribs, with the ribs partially surrounding the shoulder, and the assembly of porcine tissues inserted into a blue wax cylinder with an inner radius of 20 cm.  The pCT images used an approximate total of 180 million protons at energies of 120, 160, 185, and 203 MeV to reconstruct a volume of $250 \times 250 \times 250$ 1 mm$^3$ voxels.

The second sample was a pig's head, which we immobilized on the platform with plastic foam and tape that had four surface-mounted 4 mm CT markers (Fig.~\ref{setup}).  The pCT reconstruction used a total of approximately 150 million protons at energies of 120, 160, 200, and 220 MeV for a volume of $300 \times 250 \times 300$ 1 mm$^3$ voxels. 

We obtained x-ray CT scans of the porcine samples within 30 minutes of the pCT scans.
To keep the anatomy consistent, we acquired the x-ray CT scans with the samples in the same orientation as for the pCT scans.  We scanned the porcine girdle and ribs using a clinical vertical (horizontal-plane) CT scanner (P-ARTIS CT, P-Cure, Israel) set for 120 kVp, 325 mAs, and 0.75 mm slice thickness, and image reconstruction using the Standard B kernel. We scanned the pig's head using the same vertical CT scanner with identical settings as for the porcine girdle and ribs.  In addition, we scanned the pig's head with a clinical horizontal CT scanner (GE LightSpeed VCT64, GE Healthcare, United States) with either a high dose setting (449 mAs) or using the Automatic Exposure Control feature (AutomA) which used 49 mAs and 120 kVp and 1.25 mm slice thickness.  We used the Standard kernel for each.  We designate the scans implementing the AutomA feature as low dose scans.


We report the dose for the x-ray CT scans using the NMCPC calibrations relating the protocol settings to the expected dose in a standard CT dose index (CTDI) phantom. For the vertical CT, the CTDI was 27 mGy for the porcine shoulder and ribs, and 39 mGy for the pig's head. For the horizontal CT, the CTDI to the pig's head was 3.9 mGy for the low dose protocol and 35 mGy for the high dose protocol.

\begin{figure}
  \begin{center}
  \includegraphics[width=0.44\textwidth]{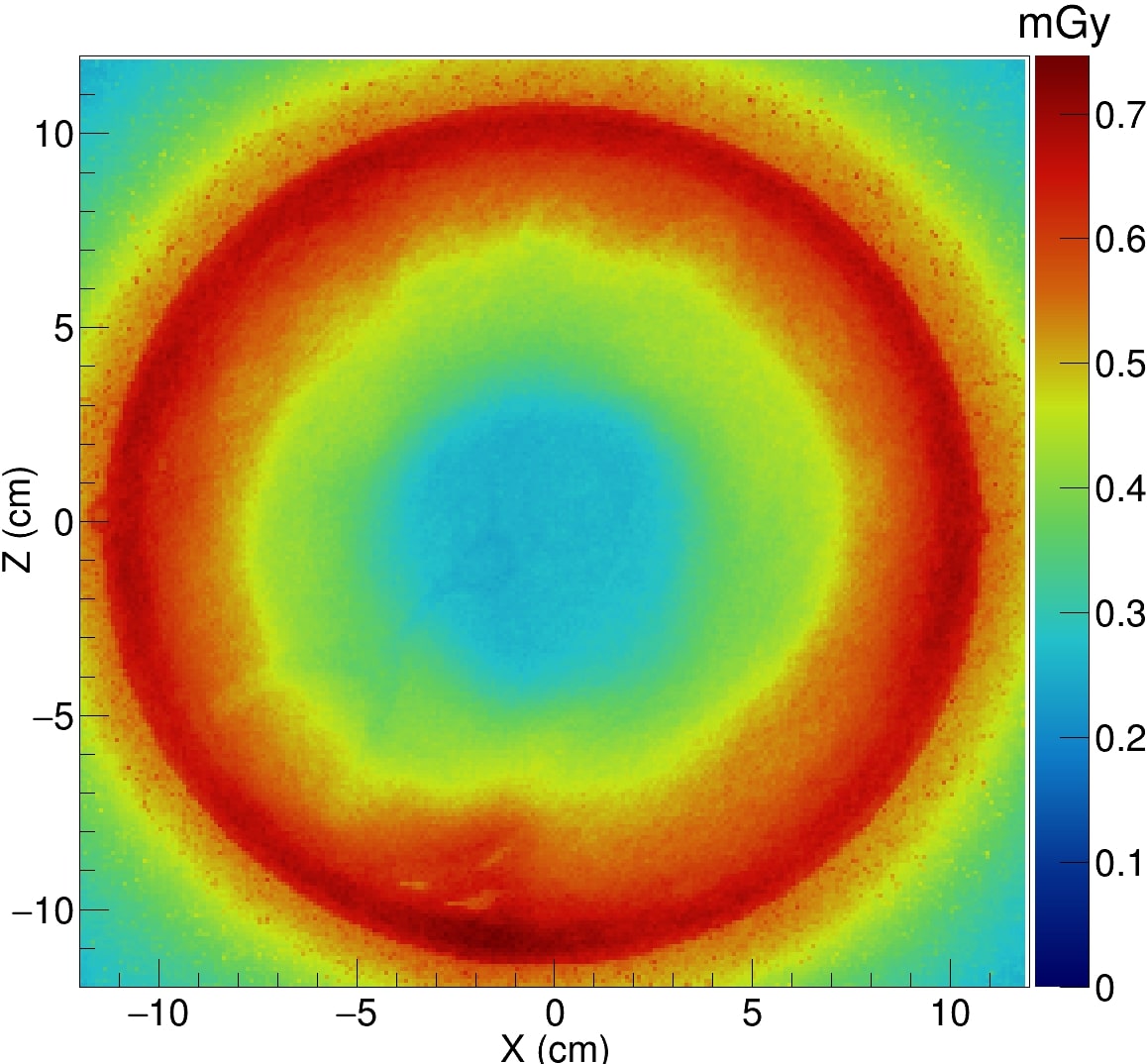}\includegraphics[width=0.44\textwidth]{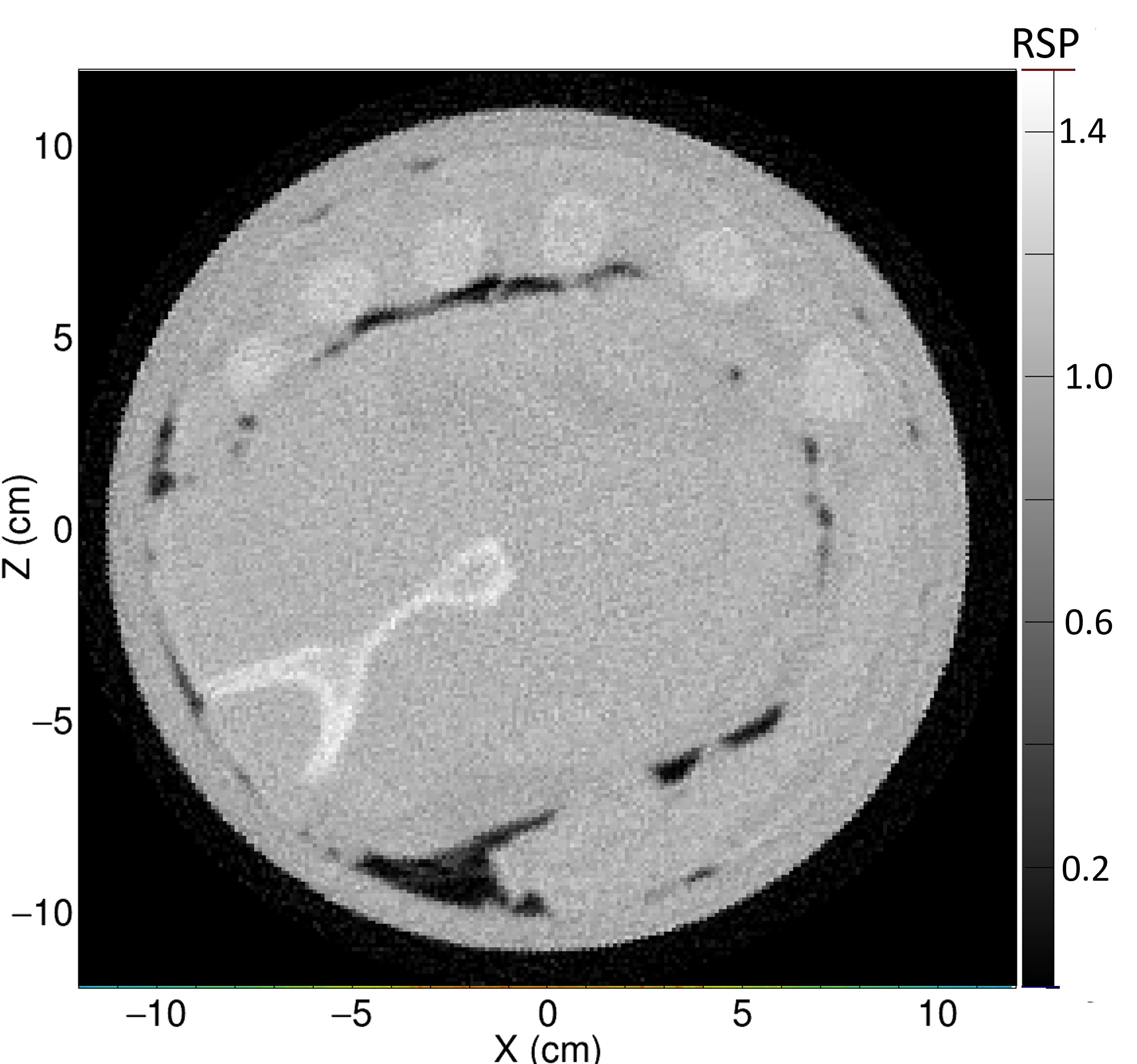}\\
  \includegraphics[width=0.44\textwidth]{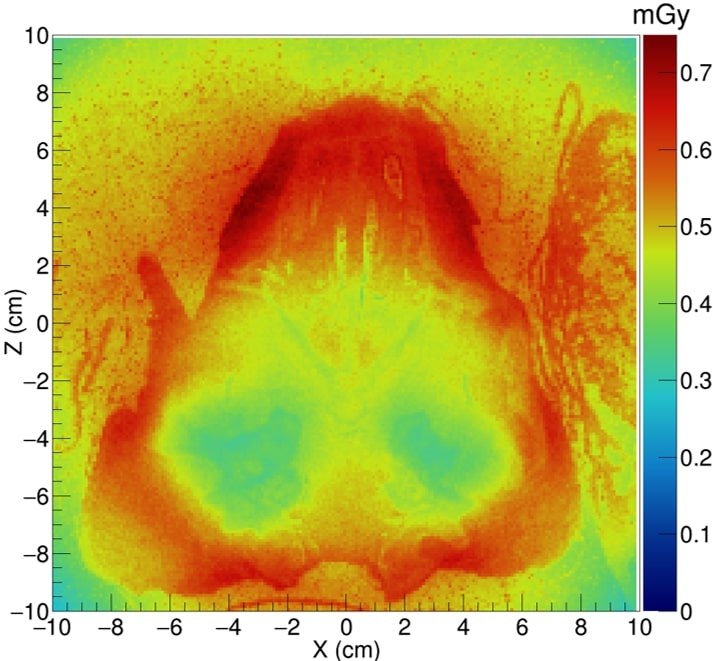}\includegraphics[width=0.44\textwidth]{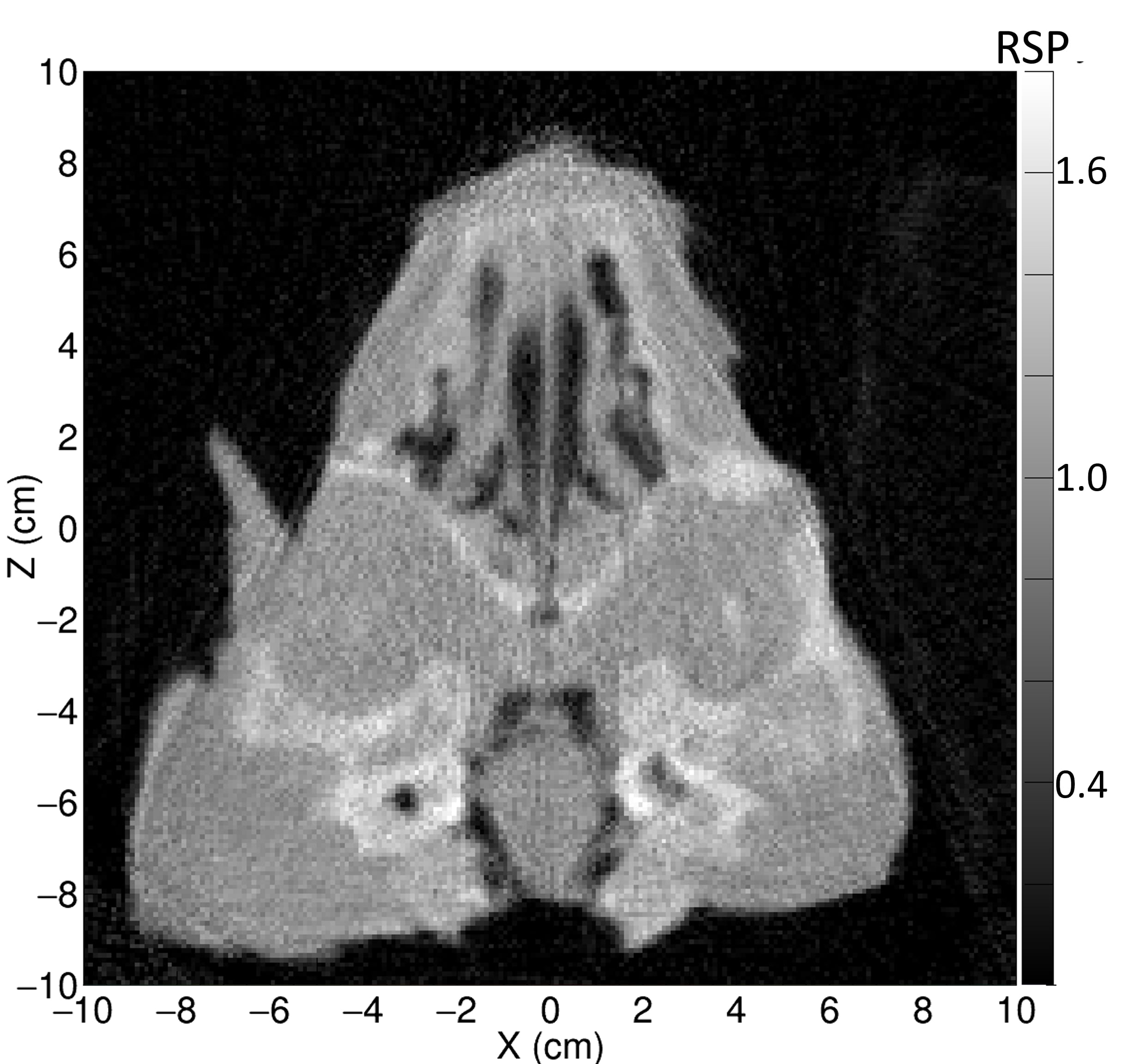}\\
  \caption{pCT dose maps for representative 2 mm slices of the pectoral girdle and ribs (top) and pig's head (bottom) for the case of optimized scan patterns, {with corresponding pCT slices on the right.}}\label{dose}
  \end{center}
\end{figure}

Since proton dose is much more dependent on density distributions than x-ray dose, we have calculated dose maps for the pCT scans, as shown in
Fig.~\ref{dose} for one representative slice for each of the porcine samples.
To obtain the data for the pCT images, we used a spatially uniform scan pattern at each angle and energy. This led to many protons not stopping in the range detector due to the residual range exceeding the thickness of the detector, while other protons stopped in the object, depending on the WET of the object for different protons. Thus,
the different proton energies contributed to different regions of the 2D projections of the object for each angle.  In future clinical use, non-uniform scan patterns will be optimized to minimize patient-specific dose while producing equivalent images. For the image reconstruction, and the dose maps shown in Fig.~\ref{dose}, we pre-selected proton events that were directed towards these regions.  
Ref.{\setcitestyle{numbers}[\cite{dejongh2020technical}]} contains additional details of the pre-selection procedure for a single angle and also shows an example of the dose difference between a uniform scan and an optimized scan. 
We used the proton fluences after this pre-selection with the TOPAS simulation program to calculate the dose maps for each angle as expected for an optimized scan pattern.
We obtained the pCT dose maps in Fig.~\ref{dose} by summing the individual projection doses over all angles.
There is large variation in the distribution of dose arising from variation in proton intensity settings at different energies.  As a result, the dose for the pig's head ranges from 0.2 mGy in regions reached only by the highest energy protons to 0.7 mGy around the edges.  In future work, equalizing the intensities at different energies will result in more uniform dose maps, and more generally, fluence modulation can optimize image noise in specific regions of interest~\cite{Dickmann_2021}.

\subsection{Image registration, contouring, and analysis}

We used the Velocity (Velocity AI 3.1.0, Velocity Medical Solutions, United States) clinical image staging program to perform a rigid registration of corresponding x-ray CT and pCT scans, resample them into the same voxel space, and define identical ROIs for each by drawing contours slice-by-slice in several segments of tissue. We started with the initial automatic registration then proceeded with manual adjustments prioritizing large bony landmarks, as well as some features of interest such as the sinus. During the manual contouring process, we ensured for each slice that the contours represented identical anatomical regions in each scan.

To convert the HU values into RSP, we used the scanner-specific stoichiometric calibration.  For each pCT ROI, we calculated the mean, standard deviation (SD), and standard error (SE). 

\subsection{pRad image acquisition and analysis}

For the pig's head, we obtained a pRad image, representing a water equivalent thickness (WET) map, in addition to the pCT data. The pRad image acquisition used a total of approximately 20 million protons of the same four energies as the pCT.

We also calculated a pRad DRR from the vertical high-dose x-ray CT scan, using the TOPAS simulation software\cite{TOPAS,TOPAS2}. 
We registered the x-ray CT to the pCT and applied the clinical Hounsfield-to-RSP conversions. The TOPAS simulation defined the isocentric reference frame and proton directions in that frame to match the beam line at NMCPC, and positioned the simulated pig's head to match the position we used for the real pRad data acquisition.  This positioning required some fine-tuning, using the surface-mounted CT markers. In order to calculate the projected WET for each pixel in the DRR, we simulated protons with energy loss effects.  In order for the DRR to represent the resolution of the original x-ray data, the proton simulations did not include multiple Coulomb scattering and the spatial resolution of the acquired pRad could therefore produce discrepancies relative to the pRad DRR in regions of rapid WET variation. We produced a WET difference map by subtracting the calculated pRad DRR from the acquired pRad.

\section{Results}

\begin{figure}
  \begin{center}
  \includegraphics[width=0.44\textwidth]{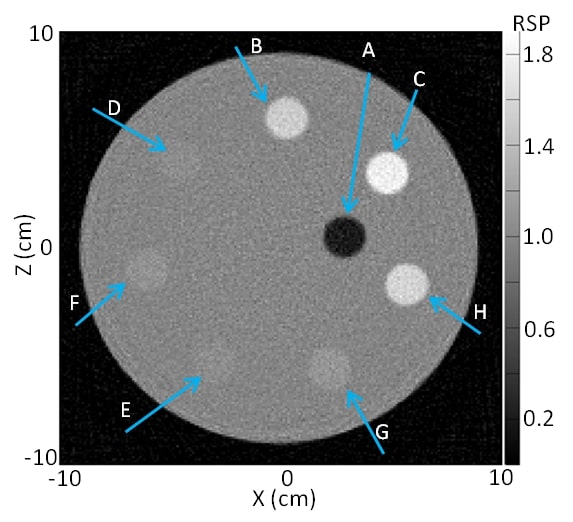}\\
  \caption{Example of a 1 mm thick pCT slice of the custom phantom with inserts of various tissue-equivalent materials with known RSP.
  Labels are: A: Sinus.
{
B: Dentin.
C: Enamel.
}
D: Brain.
E: Spinal Cord.
F: Spinal Disk.
G: Trabecular Bone.
H: Cortical Bone.}\label{george}
  \end{center}
\end{figure}

\subsection{Test of pCT accuracy with a custom cylindrical phantom}

The results in Fig.~\ref{george} and Table~\ref{table:1} demonstrate an RSP accuracy equal to or better than 1\% with the exception of a -4\% discrepancy between the pCT and the previously measured values for the sinus insert.  However, this insert has very low RSP and the absolute discrepancy is only -0.008.  The SD values show that our pCT RSP measurements have a statistical noise of typically 7\%.  

\begin{table}
\caption{Comparison of the RSP measured using pCT to the known RSP, for the tissue-equivalent inserts in the phantom in Fig.~\ref{george}.}\label{table:1}
\begin{center}
\begin{tabular}{ l|c|r|r } \hline
Insert & pCT RSP & Known & Diff. \\
& Mean\pz\pz SD\pz\pz SE(\%) & RSP & (\%)
 \\ 
\hline
 Sinus              & 0.192\pz 0.094\pz 1.5 & 0.200 & -4.0 \\
 Enamel             & 1.768\pz 0.086\pz 0.2 & 1.755 & 0.7 \\
 Dentin             & 1.504\pz 0.083\pz 0.2 & 1.495 & 0.6 \\
 Brain              & 1.043\pz 0.078\pz 0.2 & 1.040 & 0.3 \\
 Spinal Cord        & 1.046\pz 0.071\pz 0.2 & 1.040 & 0.6 \\
 Spinal Disk        & 1.079\pz 0.075\pz 0.2 & 1.070 & 0.8 \\
 Trabecular Bone    & 1.106\pz 0.079\pz 0.2 & 1.100 & 0.5 \\
 Cortical Bone      & 1.570\pz 0.084\pz 0.2 & 1.555 & 1.0 \\
\hline
\end{tabular}
\end{center}
\end{table}

\subsection{Comparison of acquired pRad with simulated pRad DRR}

 \begin{figure}
  \includegraphics[width=0.43\textwidth]{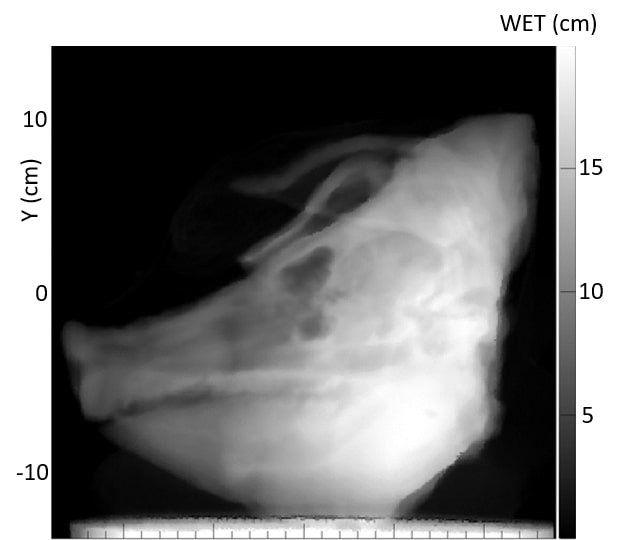}\\
  \includegraphics[width=0.43\textwidth]{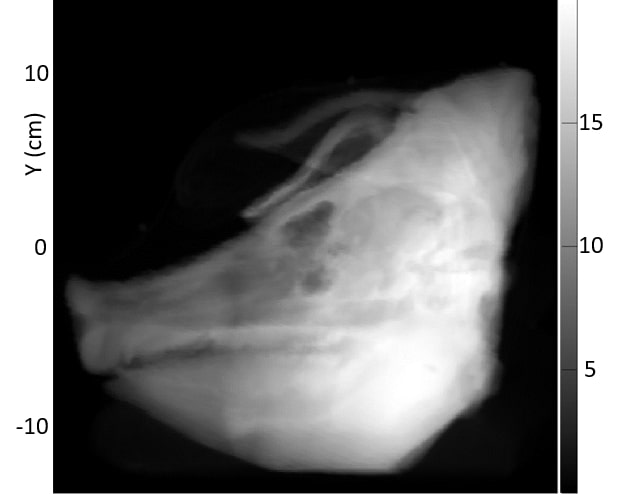}\includegraphics[width=0.44\textwidth]{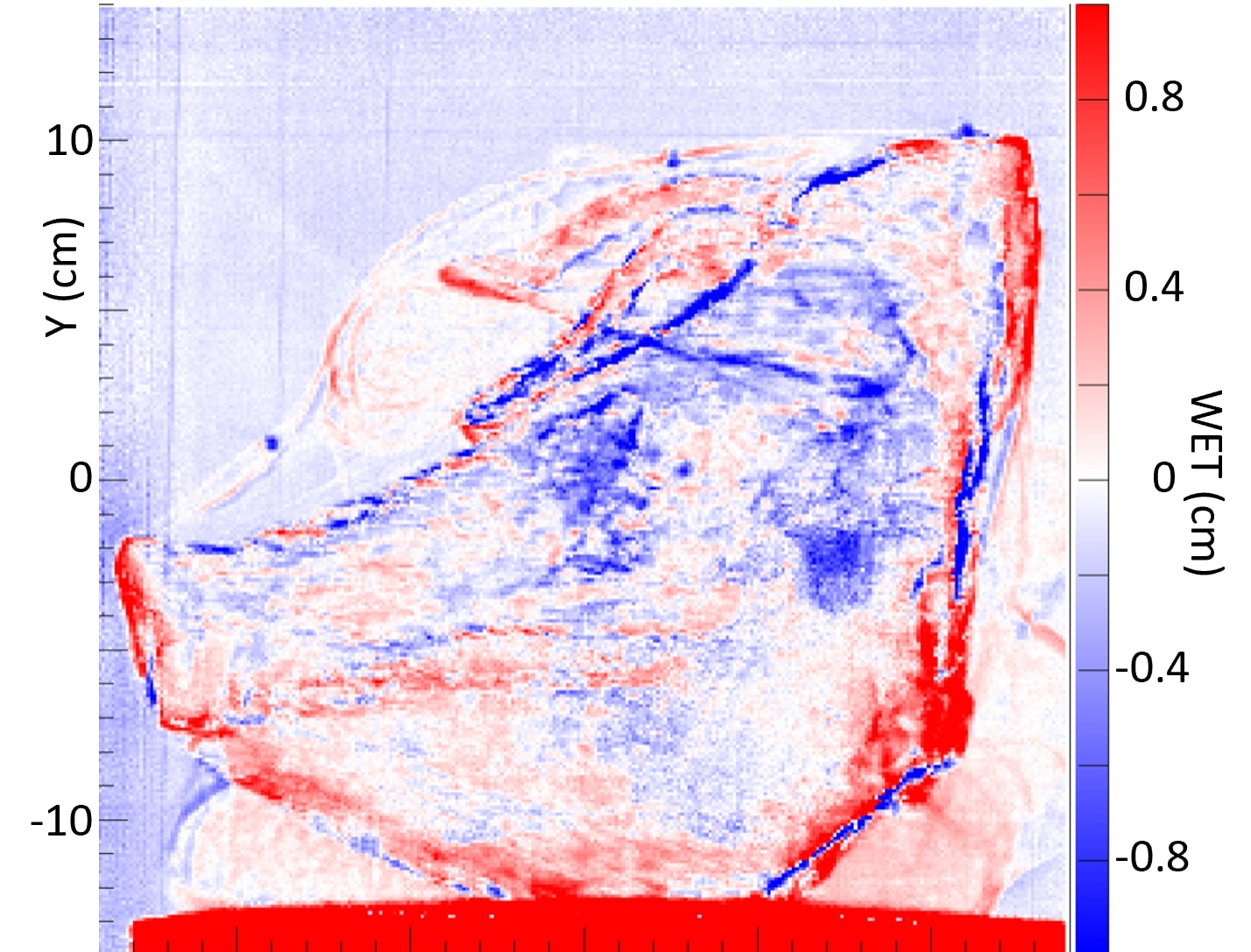}\\
  \includegraphics[width=0.43\textwidth]{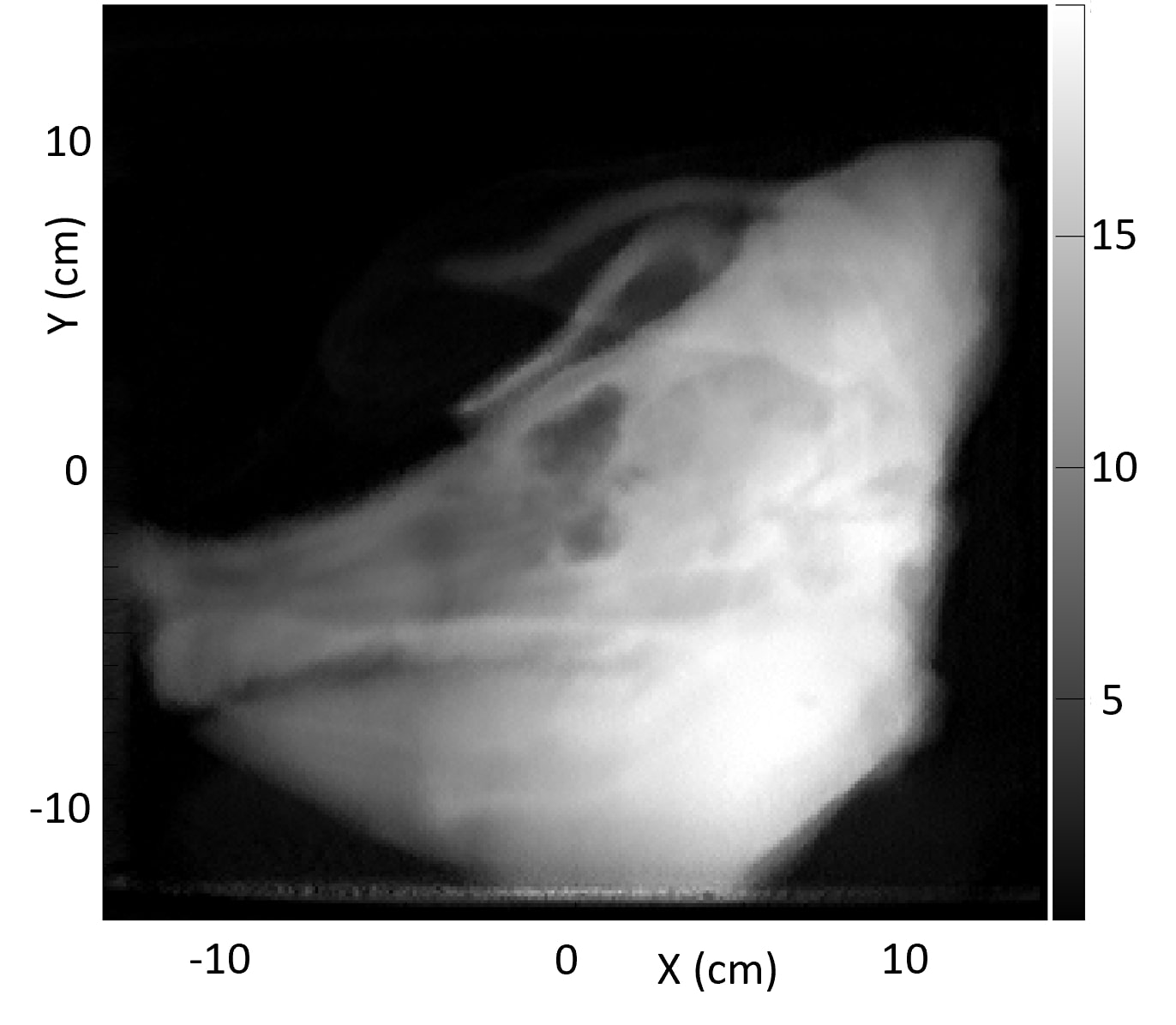}\includegraphics[width=0.44\textwidth]{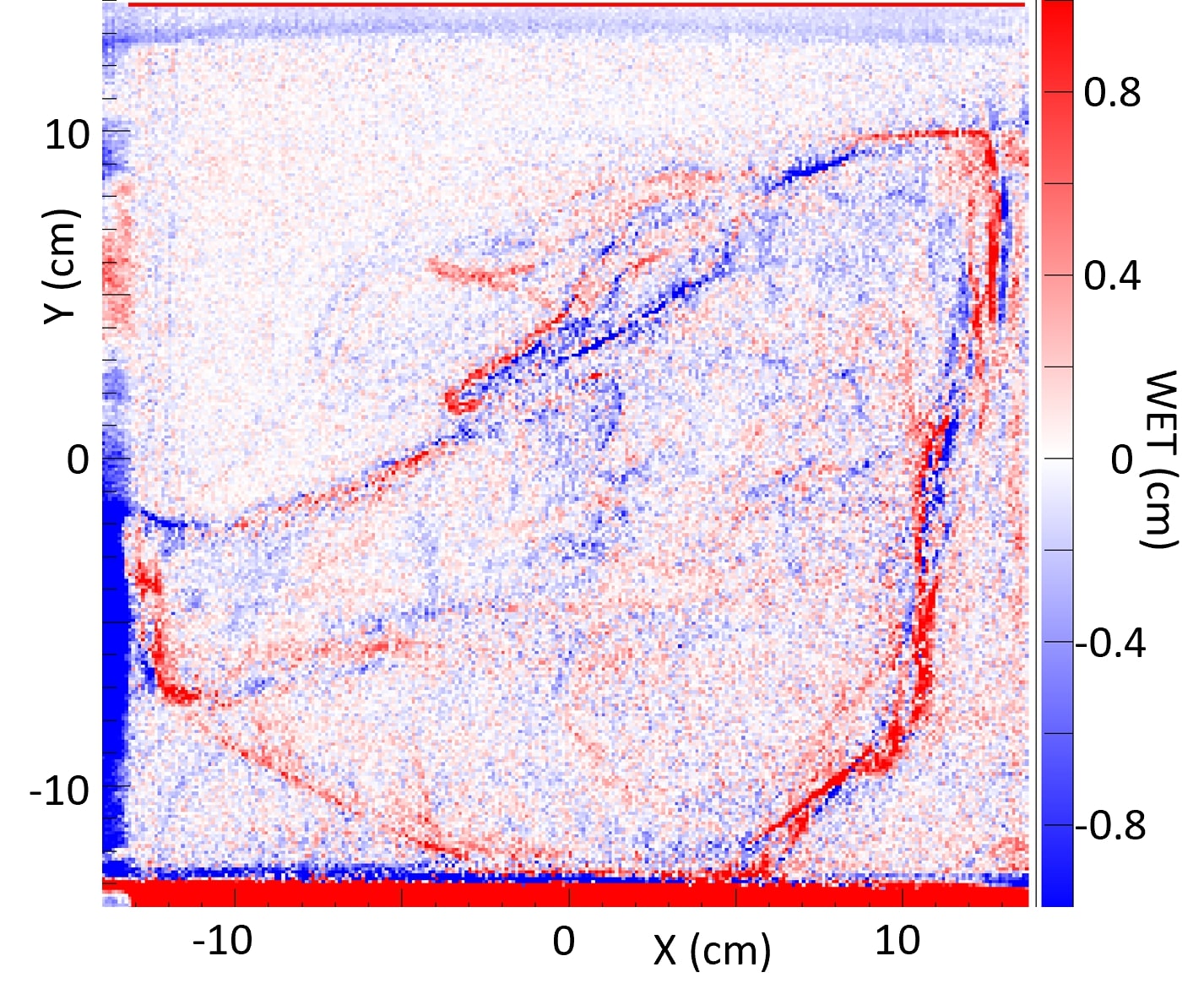}
  \caption{Top row: Acquired pRad WET image of the pig's head, used to create the difference images in the middle and bottom rows. Middle row: Simulated pRad DRR from x-ray CT (Difference from acquired pRad on right). {Bottom row: Simulated pRad DRR from pCT (Difference from acquired pRad on right).} Noteable differences with the x-ray DRR are visible in the sinus region near coordinates (0,0) and the tympanic bullae near coordinates (8,-2).}\label{pRad}
\end{figure}

Fig.~\ref{pRad} shows the acquired pRad WET image of the pig's head, along with the simulated pRad DRR from the x-ray CT with the corresponding difference image, {and a simulated pRad DRR from the pCT with the corresponding difference image.}
Relatively large differences around the edges of the head and other areas of rapid variation occur because they are sensitive to a precise alignment of the two images. The four CT markers are visible in the x-ray CT difference image (Figure 4 middle row) at coordinates near (-9,1), (3,0), (3,9), and (11,10) cm. These markers have high x-ray attenuation but do not have a correspondingly high RSP.  They are well-aligned in the difference image and appear with net negative WET values since x-ray CT overestimates the RSP of the marker materials. {In the case of the pCT difference image (Figure 4 bottom row), the marker RSPs are identical, and therefore the markers are not visible.} The slight difference of about 1 mm observed in the air around the pig's head is likely due to a calibration issue for the 120 MeV protons that travelled deep inside the scintillator. There is a faint grid structure visible in the background of the x-ray CT difference image; this is an artifact due to non-uniformities in the construction of the tracking plane.  We are currently upgrading the system, and expect to correct these non-uniformities in the next version.
 
 Most soft-tissue regions of the head, including the brain and head and neck muscles, show agreement within 1-2 mm of WET between the acquired pRad and simulated pRad DRR, $<$ 1\% in comparison to a total WET of up to 200 mm. The most notable anatomical WET differences are in the sinus region and the tympanic bullae, where the measured WET from the real pRad is 4 to 8 mm lower than the WET from the simulated pRad DRR. The sinus result is similar to that reported for a pig's head in Deffet {\it et al.} \cite{https://doi.org/10.1002/mp.14571}; however their range-probe image did not contain the tympanic bullae. {These large differences do not appear in the simulated DRR from pCT.}
 There are smaller discrepancies in bony regions including the skull and mandible.


\subsection{Comparison of pCT and x-ray CT for {porcine pectoral girdle} and ribs}

\begin{figure}
  \begin{center}
  \includegraphics[width=0.44\textwidth]{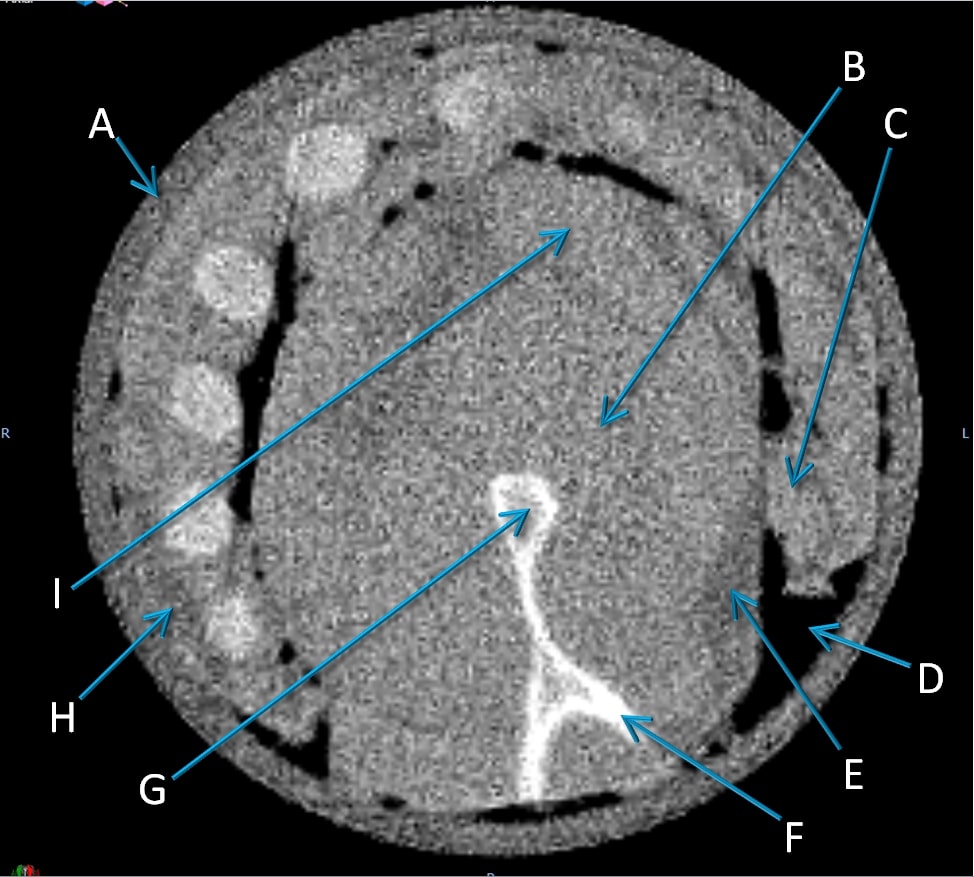}\\
  \includegraphics[width=0.44\textwidth]{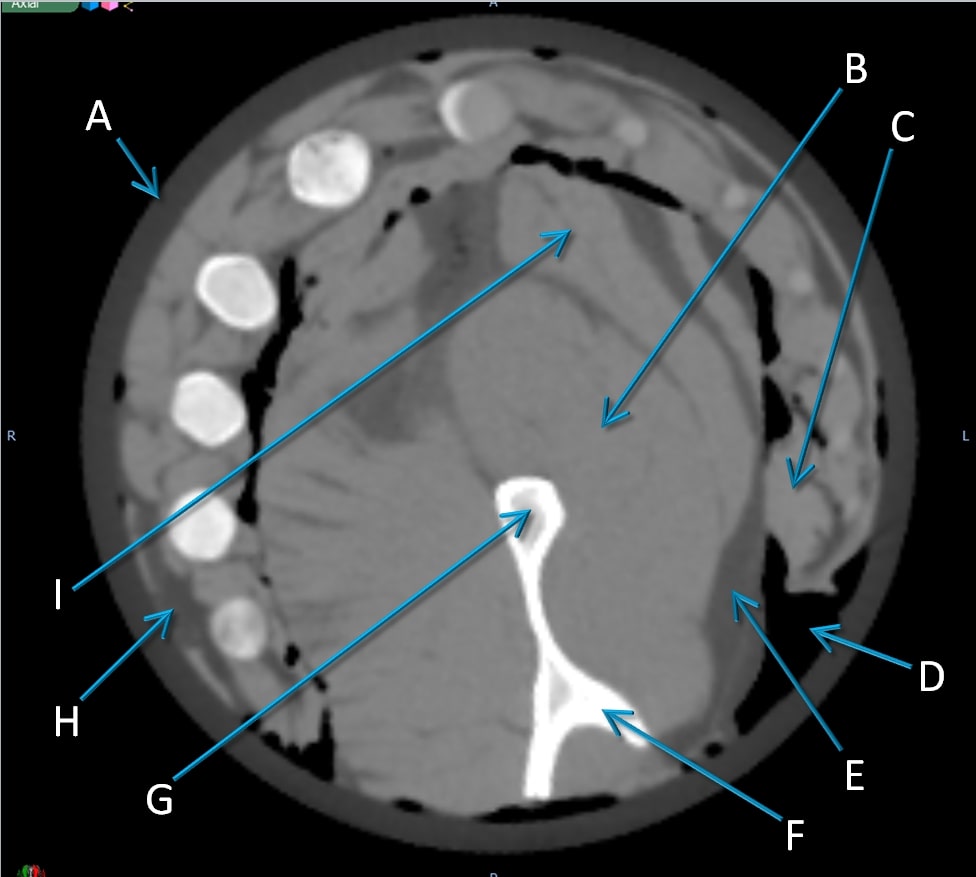}
  \caption{Examples of 1 mm thick CT slices for the {porcine pectoral girdle} and ribs, showing regions of interest.  Top: pCT.  Bottom: x-ray CT. Labels are: A: Blue Wax.
B: Muscle (Shoulder-Med).
C: Muscle (Ribs).
D: Air.
E: Adipose (Shoulder). 
F: Compact Bone.
G: Trabecular Bone (Shoulder).
H: Adipose (Ribs).
I: Muscle (Shoulder-Lat)
}\label{pork1}
  \end{center}
\end{figure}

We analyzed RSP differences between pCT and x-ray CT for 10 different regions in the porcine pectoral girdle and ribs sample, with examples of regions for one slice indicated in Fig.~\ref{pork1}. We found RSP differences of 0.6\% or less for all soft tissues, both muscle and adipose, as shown in Table~\ref{table:pork}. We see a slightly larger difference of 1.9\% in the rib trabecular bone, and a much larger difference of 6.9\% in the compact bone. The observed 4.1\% difference in the blue wax surrounding the sample is not surprising, as the Hounsfield-to-RSP conversion was not calibrated for that type of material.

Due to multiple Coulomb scattering, the pCT image appears less sharp than the x-ray CT image, as shown in the single slice comparison in Fig.~\ref{pork1}. The pCT image is still able to distinguish between muscle and fat tissue and compact and trabecular bone.

\begin{table}
\caption{Comparison of pCT and x-ray CT RSP for {porcine pectoral girdle} and ribs.}
\label{table:pork}
\begin{center}
\begin{tabular}{ l|r|c|l|r } 
\hline
\multicolumn{2}{c|}{Region \hfill Volume} & pCT RSP & x-ray CT & Diff. \\
& (cm$^3$) & Mean\pz SD\pz SE(\%) & RSP & (\%)
\\ 
\hline
 Air                    & 3.7 & 0.017\pz 0.150\pz 15 & 0.006    & 64 \\
 Adipose (Shoulder)       & 6.6 & 0.983\pz 0.086\pz 0.1 & 0.977   & 0.6 \\
 Adipose (Rib)            & 1.2 & 0.965\pz 0.054\pz 0.2 & 0.967    & -0.2 \\
 Muscle (Shoulder-Med)    & 17.5 & 1.044\pz 0.112\pz 0.1 & 1.046    & -0.2 \\
 Muscle (Shoulder-Lat)    & 25.6 & 1.043\pz 0.114\pz 0.1 & 1.043    &  0.0 \\
 Muscle (Rib)             & 5.8 & 1.051\pz 0.091\pz 0.1 & 1.046    & 0.5 \\
 Trabecular Bone (Rib)    & 1.1 & 1.120\pz 0.062\pz 0.2 & 1.141    & -1.9 \\
 Trabecular (Shoulder)     & 1.1 & 1.116\pz 0.082\pz 0.2 & 1.114    & 0.2 \\
 Compact Bone           & 0.4 & 1.467\pz 0.127\pz 0.4 & 1.568    & -6.9 \\
 Blue Wax               & 6.2 & 0.972\pz 0.114\pz 0.1 & 0.932    & 4.1 \\
\hline
\end{tabular}
\end{center}
\end{table}

\subsection{Comparison of pCT and x-ray CT for the pig's head}

\begin{figure*}
  \begin{center}
  \includegraphics[width=1.0\textwidth]{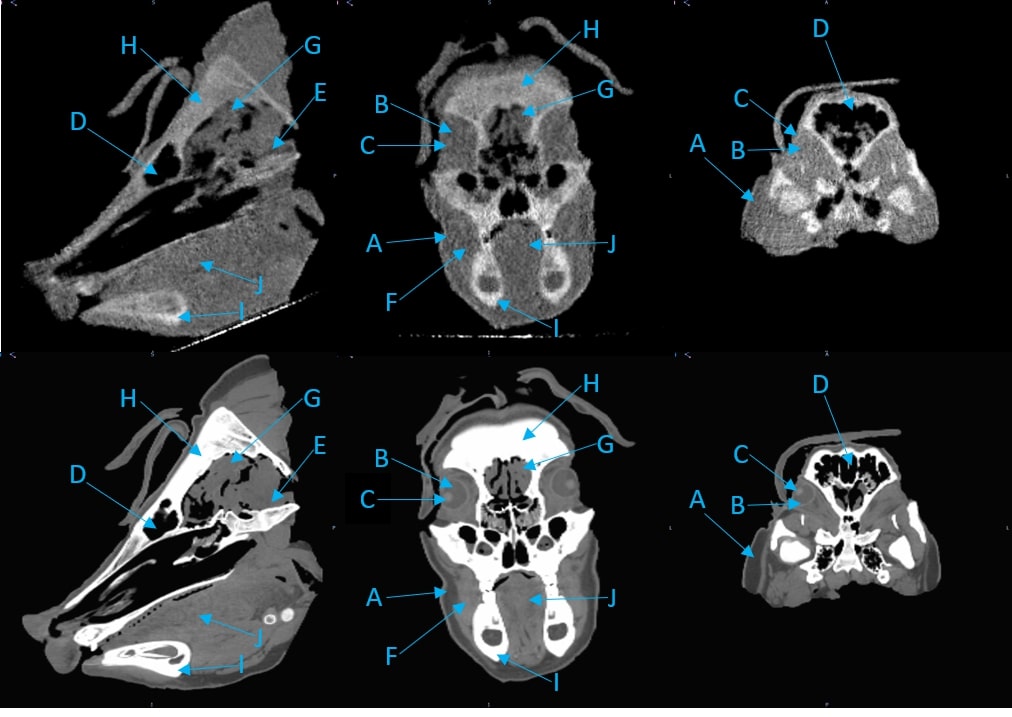}
  \caption{Examples of 1 mm thick CT slices for the Pig's head, showing regions of interest. Images on the top are for pCT, those on the bottom are for low-dose horizontal x-ray CT. Left: sagittal view. Middle:  coronal view. Right: axial view.  Labels are: A: Adipose.
B: Eye.
C: Lens.
D: Sinus.
E: Brain Stem.
F: Muscle.
G: Brain.
H: Skull.
I:~Mandible.
J: Tongue. }\label{head1}
  \end{center}
\end{figure*}

\begin{figure}
  \begin{center}
  \includegraphics[width=0.44\textwidth]{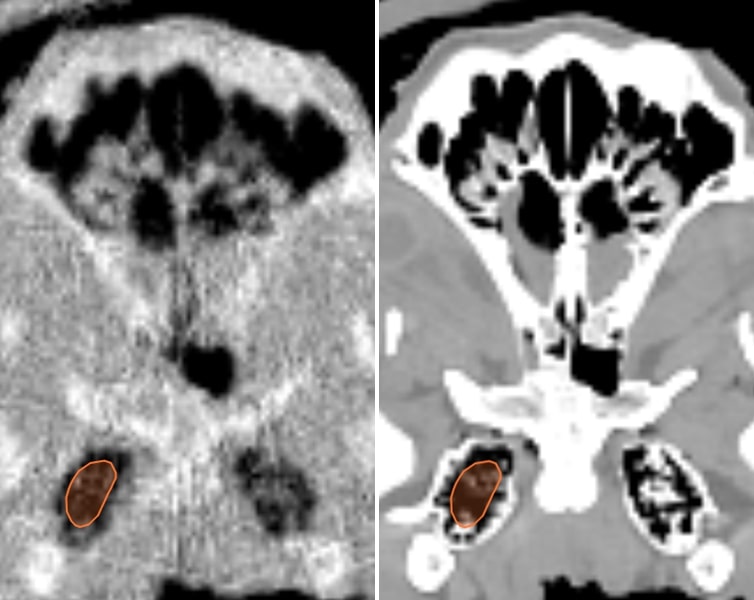}
  \caption{Tympanic bullae contours in one axial CT slice.  Left:  pCT. Right:  Low-dose horizontal x-ray CT.}\label{bullae}
  \end{center}
\end{figure}

Examples of regions for one slice each in three views of the pig's head are indicated in Fig.~\ref{head1}, and RSP comparisons between the pCT and the three x-ray CT scans are in Table~\ref{table:head}.  A close up axial view of the tympanic bullae is shown in Fig~\ref{bullae}, which illustrates that the tympanic bullae ROI encompasses a heterogeneous mixture of pneumatized cells separated by thin bony septa. The high and low dose horizontal CT RSPs are consistent within 0.5\%.  The vertical x-ray CT scan generally shows similar results, although with differences of $\sim$2\% for brain stem and skull relative to the horizontal scans.
Ignoring the RSP differences for sinus air, which are insignificant in absolute terms, the largest difference between pCT and x-ray CT is for the bullae, (-29\% to -41\%), followed by skull (-2.4\% to -4.3\%) and brain stem (-2.2\% to -4.4\%). All other RSP differences (n=24 for 8 tissues) range from -2.5\% to +2.1\% with a mean of -0.4\%. With the exceptions of tongue and lens, the differences for these eight tissues are negative or zero.



\begin{figure}
  \begin{center}
  \includegraphics[width=0.44\textwidth]{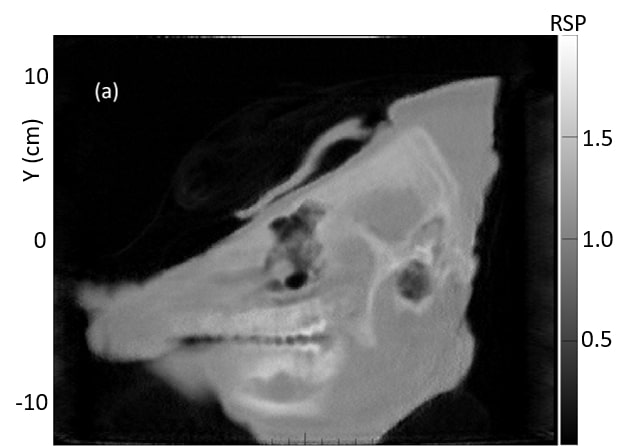}\\
  \includegraphics[width=0.44\textwidth]{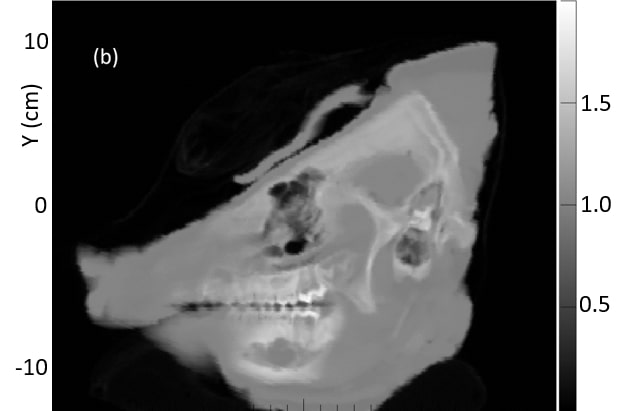}\\
  \includegraphics[width=0.44\textwidth]{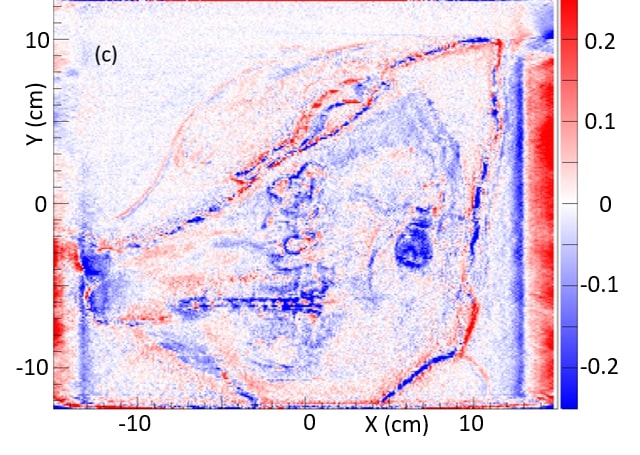}
  \caption{Ten-slice averages (a) Pig's head pCT.  (b) Pig's head vertical x-ray CT.  (c) pCT - x-ray CT difference.}\label{slices}
  \end{center}
\end{figure}

Fig.~\ref{slices} shows a comparison of the average of ten 1 mm sagittal slices from a lateral region of the pig's head for pCT and the vertical x-ray CT, in order to reduce noise. Features are slightly blurred due to their change in shape with depth. The grey scale reflects the measured RSP, whereas in Fig.~\ref{head1} and \ref{bullae} the grey scales were windowed and levelled to optimize contouring. The largest differences are in regions of heterogeneity such as the teeth and tympanic bullae.


The difference map also clearly shows the systematic discrepancy in the skull and the mandible, while soft tissue regions such as brain and muscle regions show closer agreement. The difference in the tip of the snout is caused by an incomplete set of pCT data through that area. The pencil beam scans used for the pCT data set only covered a $25 \times 25$ cm$^2$ area, so the tip of the snout was not quite covered at lateral angles. 

\begin{table}
\begin{adjustwidth}{-1cm}{}
\caption{Comparison of pCT and x-ray CTs RSP for the pig's head. Differences are between the pCT and respective x-ray scans.}
\label{table:head}
\begin{center}
\begin{tabular}{ l|r|c|l|r|l|r|l|r } 
\hline
\multicolumn{2}{c|}{Region \hfill Volume} & pCT RSP & Hor CT$^a$ \ & Diff & Hor CT$^b$ & Diff & Vert CT & Diff \\
& (cm$^3$) & Mean\pz SD SE(\%) & RSP & (\%) & RSP & (\%) & RSP & (\%) 
 \\ 
\hline
 Bullae                 & 0.8 & 0.491\pz0.24\pz 1.7 & 0.684  & -39.3 & 0.690 & -40.5 & 0.634 & -29.1 \\
 Adipose                & 3.7 & 0.950\pz0.14\pz 0.2 & 0.961  & -1.2 & 0.962   & -1.3 & 0.954  & -0.4\\
 Muscle                 & 2.0 & 1.033\pz0.16\pz 0.3 & 1.058  & -2.4 & 1.059   & -2.5 & 1.052  & -1.8\\
 Tongue                 & 9.4 & 1.047\pz0.23\pz 0.2 & 1.035  & 1.1  & 1.036   & 1.1  & 1.031  & 1.5\\
 Brain Stem             & 0.7 & 0.994\pz0.16\pz 0.6 & 1.038  & -4.4 & 1.038   & -4.4 & 1.016  & -2.2 \\
 Brain                  & 2.5 & 1.025\pz0.16\pz 0.3 & 1.037  & -1.2 & 1.039   & -1.4 & 1.031  & -0.6\\
 Lens                   & 0.1 & 1.099\pz0.12\pz 1.6 & 1.078  & 1.9  & 1.080   & 1.7  & 1.076  & 2.1\\
 Eye Left               & 0.5 & 1.015\pz0.13\pz 0.5 & 1.015  & 0.0  & 1.017   & -0.2 & 1.018  & -0.3\\
 Eye Right              & 0.8 & 1.011\pz0.15\pz 0.5 & 1.021  & -1.0 & 1.021   & -1.0 & 1.014  & -0.3\\
 Skull             & 0.5 & 1.266\pz0.12\pz 0.4 & 1.297  & -2.4 & 1.303   & -2.9 & 1.320  & -4.3\\
 Mandible               & 0.5 & 1.540\pz0.16\pz 0.5 & 1.559  & -1.2 & 1.565   & -1.6 & 1.562  & -1.4\\
 Sinus Air              & 0.1 & 0.067\pz0.12 17\pz & 0.057  &  15\pz & 0.058   &  13\pz    & 0.039  &  42\pz  \\
\hline
\end{tabular}
\end{center}
\end{adjustwidth}
$^a$ Low dose protocol\\
$^b$ High dose protocol
\end{table}

\section{Discussion}

We have obtained several detailed pCT images of complex porcine structures using a prototype pCT scanner. Our proton WEPL calibration procedure ensures a direct and accurate RSP measurement, as verified by the results in Table~\ref{table:1}. While we expect the accuracy of our pCT measurements to apply to any object, we have planned future proton range measurements for  direct confirmation of the RSP accuracy.  For example, a small radiochromic film stack could be inserted inside a porcine sample.  A low-dose planning pCT image of the sample with the film stack would be obtained to measure the difference between planned and observed proton range for a single uniform dose treatment beam\cite{Christina}.  

For the reconstruction of the pCT images, we did not apply any filtering to suppress noise. This, combined with low proton fluences and dose ($<$ 1 mGy range), led to relatively noisy images compared to the corresponding x-ray CT images.
Upgrades to our data acquisition will enable automatic scans and routine acquisition of data sets with more protons and reduced image noise. Studies are in progress to further analyze the noise properties of the pCT image, and its impact on range uncertainty.

To our knowledge, this study is the first comparison of pCT and x-ray CT images of the same porcine samples including  direct RSP measurements, high spatial resolution, and three-dimensional ROI analysis.
Our results confirm, with several tissue-specific measurements in intact porcine structures, several recent reports~\cite{https://doi.org/10.1002/mp.14571,Meijers_2020,Meijers_2020_lungs,baer-tissue} indicating that it may be possible to reduce the standard uncertainty margins used in proton therapy. In the case of the porcine shoulder and ribs, the agreement between pCT and x-ray CT for soft tissues is within 0.6\%.  In the case of the more complex pig's head, agreement for soft tissues is within 2.5\% except for the brain stem, for which there is a 4.4\% difference relative to the horizontal CT scanner and a 2.2\% difference relative to the vertical CT scanner. These differences could potentially be due to systematic uncertainties caused by small registration offsets in heterogeneous regions, or artifacts near bony interfaces.

Our results show larger discrepancies in bony regions, with differences up to -7\%, and the pattern of this difference for bones in the pig's head are visible in the pCT -- x-ray CT difference image of~\ref{slices}. The largest differences are in cavitated regions such the sinus and tympanic bullae, with differences up to 40\%.  A previous study using a range probe approach~\cite{https://doi.org/10.1002/mp.14571} also observed large discrepancies in the sinus region of a pig's head, indicating that these results may apply in general.

Our high-quality images and results indicate the potential for pCT imaging to be used as a low-dose clinical treatment planning modality with reduced range uncertainties. DECT systems can also reduce range uncertainties, by measuring two sets of CT numbers at different X-ray energies. 
A first direct experimental comparison of pCT and DECT~\cite{pctanddect}, using a pCT system with a multi-stage range detector~\cite{LomaLinda} and a set of phantoms with tissue-mimicking inserts, found an accuracy for both better than 1\%. The pCT accuracy was limited by artifacts arising from the segmentation of the multi-stage range detector, a promising conclusion for the monolithic ProtonVDA range detector.

{DECT produces higher dose to the patient than pCT}, and the images will suffer from artifacts in the presence of metallic implants. However, DECT may be able to provide imaging in cases where there is not enough proton energy for a full pCT image. The two imaging modalities are therefore in some ways complementary, with proton imaging potentially providing direct RSP measurements useful for calibration checks of DECT in a wide variety of circumstances including with human patients.

In addition to its use for pCT, a proton imaging system could potentially be used to acquire one or more pRad images with an efficient workflow before each treatment.
Comparisons between pRad images and DRRs from the planning CT, as for the example in Fig.~\ref{pRad}, could help provide assurance that treatment plans using reduced margins are safe. Furthermore, it may be possible to use one or more pRad images to detect inaccuracies in the planning CT. For example, several recent studies have explored methods to use pRad images to produce patient-specific x-ray CT calibrations~\cite{Hansen_2014,Doolan_2015,Collins_Fekete_2017,Krah_2019,gianoli} and thus more accurate RSP maps.  

Proton beam therapy can be useful for radiotherapeutic management of many head and neck and intracranial malignancies. 
Clinical data leads us to believe that there is room for improvement with proton beam therapy and reduced margins, with the possibility of safely escalating dose to improve local control while maintaining the lower rates of toxicity imparted by proton beam therapy. 
{Treatment planning for intensity modulated proton therapy (IMPT) for head and neck cancer patients currently uses clinical tumor volume (CTV) based robust optimization}\cite{IMPT}. {This approach has replaced planned tumor volume (PTV) based planning that accounts for range uncertainties by expanding CTV to PTV via additional margins.  The robust CTV-based approach optimizes dose distributions under assumed setup and range uncertainties. Incorporating pCT with its reduced range uncertainty into this planning scheme will be straightforward. The range uncertainty reduction achievable with pCT was previously studied}\cite{accuracy} {using TOPAS with a digitized pediatric patient head CT model. The pCT scan with a realistic scanner model, comparable to the scanner used in this work, resulted in range errors typically under 1 mm, except for beam directions parallel to a bone soft-tissue boundary.}

Among the classic head and neck examples for which proton beam radiation therapy appears especially useful are esthesioneuroblastoma (olfactory neuroblastoma), sinonasal undifferentiated carcinoma (SNUC) and some challenging nasopharyngeal carcinoma cases. In such cases, proton beam therapy can offer additional sparing of critical adjacent sensitive structures, such as the globe, optic nerves, optic chiasm and brain  compared to conventional photon-based external beam radiation therapy~\cite{Nichols}. This is needed since, for instance, the conventional radiation therapy literature on esthesioneuroblastoma reports severe ocular radiation injury leading to a poor or nonfunctioning eye at rates ranging from 8\% to 24\%~\cite{Simon,Dulguerov,Hwang}.

As an example of a specific successful case, Thekkedath \textit{et al.} reported the case of a 31-year-old Bosnian male who presented with headaches, facial pain, and epistaxis who was found to have a very large SNUC. The patient experienced near complete resolution of his stage IVB (T4b N0 M0) SNUC after treatment with combination chemotherapy (cisplatin and gemcitabine followed by weekly cisplatin) in combination with proton therapy~\cite{Thek}.  SNUC is a difficult cancer to conquer and this case illustrates the potential of proton beam therapy. Refinements in dose delivery will help ensure similar outcomes in a larger fraction of cases.

\section{Conclusions}

In a comparison of proton stopping power as measured directly with proton CT to values obtained from x-ray CT scans using clinical scanner-specific HU to RSP calibrations,
we found agreement within 1\% to 2\% for most soft tissues, and discrepancies of up to 7\% for compact bone. We observed the largest discrepancies, up to 40\%, for cavitated regions with mixed content of air, soft tissue, and bone, such as sinus cavities or tympanic bullae. If these results apply in general, it may be possible to substantially reduce uncertainty margins for many treatment plans, while avoiding regions with higher uncertainties.  Our comparison of proton radiography data with a digitally reconstructed radiograph from the x-ray CT demonstrates the potential of this modality for daily pre-treatment range verification. Our images and findings from a clinically realistic proton CT scanner demonstrate the potential for proton CT to be used for low-dose treatment planning with reduced margins.

\section*{Acknowledgments}

Research reported in this publication was supported by the National Cancer Institute of the National Institutes of Health under award number R44CA243939.

The authors thank Aditya Panchal of Amita Health for assistance with converting pCT images to DICOM format. We thank the staff of NMCPC for assistance with and use of the treatment room and x-ray CT scanner.  We thank the staff of Ion Beam Applications for assistance with proton beam operations, including the work of Nick Detrich to produce custom pencil beam scanning patterns.  We thank
Igor Polnyi of ProtonVDA for his help assembling the detector and operations during data collection.

\section*{Conflict of Interest Statement}

The authors have intellectual property rights to the innovations described in this paper. Don F. DeJongh and Victor Rykalin are co-owners of ProtonVDA LLC.

\clearpage


\section*{References}

\vspace*{-20mm}





\bibliography{./porcine}      



\bibliographystyle{./medphy.bst}    


\end{document}